\documentstyle[12pt]{article}
\begin{document}
\title{Escape of Superheated Upsilons from the Quark Gluon Plasma}

\author{Petre Golumbeanu\thanks{petrego@suhep.phy.syr.edu} $ $ and Carl
Rosenzweig\thanks{rosez@suhep.phy.syr.edu}  \\ Department of Physics, Syracuse University}
\date{\today}
\maketitle

\begin{abstract}
The properties of heavy quark systems change if they are placed in a medium other than the low energy vacuum. In a hot Quark Gluon Plasma $J/\Psi$ particles will melt and not exist as resonant states. $\Upsilon$'s, however, because of their smaller size and the dominance of the Coulomb potential, will still form as $Q\bar Q$ bound states but their properties will shift. In particular the $\Upsilon$ mass may be shifted upward by over 150 MeV. If such excited states manage to escape from the plasma as a superheated particle, they may serve as a diagnostic of the plasma in which they originated. We propose that such a scenario is possible and that hot $\Upsilon$'s will form and escape, thereby providing us with crucial information about the Quark Gluon Plasma.
\end{abstract}

\newpage 

\section{\bf Introduction}
As the hunt for the Quark Gluon Plasma (QGP) heats up, the search for a clean 
diagnostic signal becomes pressing.  The properties of heavy Quarkonium 
have figured prominently, and promisingly, as a signal for the formation of 
the QGP.  Over a decade ago the melting of the $J/\Psi$ particle in a  plasma was suggested $\cite{1}$ as a signal for QGP.  However, the subsequent observation of $J/\Psi$ suppression in heavy ion collisions was the result of nuclear absorption rather than melting [for a review of the field, see \cite{2}].  More recent observations \cite {3}are indicative of a melting effect \cite{4,5,6} although, again, the interpretation is not unambiguous \cite{7,8,9}.
 
The advent of RHIC and LHC may give us the opportunity to clearly see this 
melting effect. The higher energies available to collide larger nuclei will 
produce hotter, more dense material. In addition we expect production of 
heavy Upsilons [see, e.g., \cite{10}]. In this note we propose that the properties of these heavy Quarkonium may expose the existence of the QGP and help in exploring its properties.

The original suggestion for quarkonium melting in QGP was based on the idea 
that the confining inter quark potential would be screened in the plasma and 
that sufficient screening would liberate the quarks. While recent discussions 
of $J/\Psi$ suppression rely on a more microscope description of gluon quarkonium interactions, the original picture is more immediately transparent and 
suggestive of further effects.  Our discussion will be framed in the context 
of a macroscopic picture of the QGP, and its inherent screening properties. 

Fragile quarkonium such as $J/\Psi$, are readily disrupted by
screening. The more robustly bound $\Upsilon$ will maintain its
integrity as a bound state in the plasma but will have its properties,
such as its binding energy and hence its mass, shifted.  Since the
screened $\Upsilon$ will still be small, comparable in size to the
unscreened $J/\Psi$, it will not be highly interactive with the hot
hadronic gas that will eventually replace the cooling plasma. It thus has a reasonable chance of escaping relatively unscathed from its fiery surroundings and 
decaying in a detector with a mass different from its unscreened, normal, 
brethren.  This mass shift will be a noticeable signal, and possible 
diagnostic of the Quark Gluon Plasma.  The remainder of this note will be a 
justification of the above scenario.

\section{\bf Hot Upsilon}

$\bf{2.1 \quad The \quad Screened \quad Cornell \quad Potential}$

\vspace{0.2cm}

Debye screening in a conventional plasma is well established.  Analogous 
effects are expected in the QGP.  A heavy Q$\bar Q$ in the medium of the QGP will 
still experience an attractive potential but both the short distance, Coloumb, 
and the long distance, linear, pieces of the potential will be screened.  If we 
assume the zero temperature, zero baryon density, interquark potential is 
described by the phenomenological addition of the long range confining 
potential and short range Coulomb potential we have the Cornell potential

\begin{equation}
V=\sigma r -{\alpha\over r}
\label{first}
\end{equation}

We expect that the Coulomb, perturbative piece will undergo Debye screenning and behave like 

\begin{equation}
-{\alpha\over r}e^{-\mu r},
\end{equation}                                                                                                                                         
with $\mu (T)$, the screening mass, an increasing function of the
temperature $T$. The Debye screening radius is $1/\mu$ .
The linear part of the potential should also be screened and we adopt the suggestion of \cite {11}

\begin{equation}
{\sigma\over \mu}(1-e^{-\mu r}),
\end{equation} 
                                                                               leading to the screened QGP potential

\begin{equation}
V(r,\mu )={\sigma\over \mu}(1-e^{-\mu r})-{\alpha\over r}e^{-\mu r}
\end{equation}

Evidently the confining character of the potential is gone, just as we anticipated.  
However, we still have an attractive potential which can bind a $Q\bar Q$ pair into quarkonium if the quarks are sufficiently massive.  
It is a dynamical question which states will bind and which will not. Even for moderately high values of T, the screened potential (4) will  bind $b\bar b$ into a bound  "hot" $\Upsilon$ with properties different from the properties, determined by (1), of the $\Upsilon$ at $T=0$. This contrasts with the case of $J/\Psi$  which readily loses its binding and melts.  This was the initial motivation for the proposal that $J/\Psi$ suppression would be a signal for the QGP.  The fact that a modified $\Upsilon$ may survive offers the opportunity to study some properties of QGP in more detail. 

In order to study the modifications introduced by screening we need an
acceptable $T=0$ potential. By now this is a routine task and has been
done numerous times. We use here the values presented in \cite{12}
(based on a fit to the spin-averaged energy levels of $\Upsilon$ family):

\begin{equation}
\alpha=0.470,\quad \sigma=0.186 Gev^{2},\quad m_{b}=4.753 Gev  
\end{equation}

We now ask what happens to the bound states when the screening turns on as $T$ increases. In our simple model all the $T$ dependence is contained in $\mu$.  

\vspace{0.2cm}

$\bf{2.2 \quad The \quad T-dependence \quad of \quad \mu}$

\vspace{0.2cm}

The $T$-dependence of the Debye mass is a subject of intense ongoing 
theoretical research [13-18]. A survey of the literature
leads to the conclusion that the leading-order perturbative result
\cite{13}  

\begin{equation}
{\mu^{LO}(T)/T}=(N_c/3+N_f/6)^{1/2}\cdot g(T) 
\end{equation}
consistently underestimates the lattice results for $\mu (T)/T$. 
Significant non-perturbative contributions are present up to very large
temperatures. (In eq.(6) $N_c$ is the number of colors,
$N_f$ is the number of light quark flavours involved and $g(T)$ is the $T$-running coupling constant.)

We are interested in the behaviour of the Debye
screening mass at temperatures ranging from one $T_c$ to a few (4-5)
$T_c$ ($T_C$ is the deconfinement transition temperature). 
In this range of temperatures there are lattice results for both SU(2)
\cite{14,15} and SU(3) \cite{16} pure gluonic theories. 

In the case of SU(2) both \cite{14} and \cite{15} report a practically linear
$T$-dependence of $\mu$, albeit with different slopes: $\mu(T)\simeq
2 \cdot T$ from \cite{14} and  $\mu(T)\simeq 2.8 \cdot T$ from \cite{15}. For the SU(3) case, the result reported in \cite{16} is
$\mu(T)\simeq 6.3 \cdot T$. 

Results including dynamical quarks are not yet available from lattice
measurements. Nevertheless, model calculations based on the method of
dimensional reduction of high $T$ QCD (see for instance \cite{17},\cite{18})
offer some guidance for the dependence of the screening mass on the
number of quark flavors. Ref. \cite{17} calculates the value of
the ( electric) screening mass at $T=2T_c$, for SU(3) and three
flavors, to be $\mu(T)\simeq 6.6 \cdot T$. (The same work offers the screening
mass for pure SU(3) at $T=2T_c$ namely  $\mu(T)\simeq 4.6 \cdot T$). 

In principle the lattice measurements should be the most reliable ones 
as they use a fully non-perturbative method. Therefore we use here the
previously mentioned linear $T$-dependence of $\mu$ allowing a range
for the proportionality constant: $\mu(T)\simeq (2.5-6)\cdot T$. As a
working case we choose

\begin{equation}
\mu(T)=4 \cdot T
\end{equation}

\vspace{0.2cm}

$\bf{2.3 \quad The \quad Hot \quad Upsilon}$

\vspace{0.2cm}

We are most interested in the mass and the size of the possible hot bound states.  The mass shift will serve as a marker for the unusual environment in which the bound state forms while the size will be a useful guide to the ability of the shifted, superheated state to slip through the hot hadronic environment and escape.  

We solved the Schrodinger equation numerically with the potential
given by eqs.(4),(5). Table 1 shows the masses and sizes of
$\Upsilon (1S)$ states for several values of the screening mass $\mu$.

\begin{table}[h]
\begin{tabular}{|r|r|r|}
\hline
$\mu$(Gev) & Mass(Gev) & radius(fm) \\\hline
0.00 & 9.454 & 0.23   \\\hline 
0.36 & 9.555 & 0.26   \\\hline
0.72 & 9.615 & 0.31   \\\hline
1.08 & 9.639 & 0.46   \\\hline
1.40 & 9.635 & 1.19   \\\hline
\end{tabular}
\caption{Mass and radius of $\Upsilon(1S)$ for different screening masses.}
\end{table}

The mass of the $\Upsilon (1S)$ is seen to increase  as $\mu$ (and $T$) increases from $0$ to $1 GeV$.  The size of the state also increases but remains relatively small.  

For instance, at $\mu=0.72GeV$, corresponding to $T=T_{c}=180MeV$, the
hot $\Upsilon (1S)$ has a mass of $9.615GeV$ ($159MeV$ greater than
the $T=0$ mass) and a size of $0.31fm$, significantly smaller than
that of a cold $J/\Psi$ ($\simeq 0.45fm$, \cite{11}).

Beyond $\mu \simeq 1.50GeV$ the potential of eq.(4) does not allow bound states, i.e. the states ``melt''.  

The mass shift is easy to understand.  The $b\bar b$ quarks in the
$\Upsilon$ primarily feel the influence of the Coulomb part of the
potential.  They are so close together that they are hardly aware of
the confining part and thus barely notice its disappearance at high
$T$.  It is this insensitivity to the confining potential that is
responsible for the survival of the $\Upsilon$ at high $T$.  If the
Coulomb potential is screened, its negative contribution to the bound
state energy is weakened, increasing the $\Upsilon$'s mass and
radius. We note here that these (relatively) small changes in the
bound state properties will not affect significantly the decay
width. The lifetime of the hot state exceeds, by large factors, any
strong interaction timescale.

If this superheated state escapes and can be observed it will
indicate the formation of a macroscopic medium with a screened
potential.  The most optimistic scenario would allow this shifted mass
to serve as a thermometer of the original QGP. Of course, at very high
$T$ even $\Upsilon(1S)$ will melt, but, as we shall argue, in the
temperature range of RHIC we expect mass-shifted $\Upsilon$s to form.  

\vspace{0.3cm}

\section{\bf Fate of Hot Upsilons in Heavy Ion Collisions}

\vspace{0.4cm}

$\bf{3.1 \quad Hot \quad States\quad and \quad P_T \quad Dependence}$

\vspace{0.3cm}                                                                              

 We begin by looking at a heavy ion collision  which produces a $Q\bar Q$ system at $t=0$, at rest in the lab (CM) frame.  After a time $t_0$, a large (effectively infinite) QGP, at temperature $T_0$, forms.  The QGP will begin to cool.  We assume the cooling rate is related to the expansion time as given by longitudinal isentropic expansion equation \cite{19}

\begin{equation}
 s(t_0)t_0  = s(t)t                                                            \end{equation}                   
where s(t) denotes the entropy density at time t.

 There are three temperatures that are of significance:  

\vspace{0.3cm}

a) $T_0$ is the initial temperature of the plasma.
 
b) $T_c$ is the transition temperature above which  a QGP can
form. Recent results from lattice computations suggest $T_c\simeq
170-190MeV$  \cite{20}. In this paper we will use the value $T_c=180MeV$.  
Below $T_c$ we have a hot hadron gas.

c) We define $T_m$ as the melting temperature above which bound
 $Q\bar Q$ states can no longer form. In our model it is the
 temperature for which the potential (4) no longer supports bound
 states. For the $\Upsilon (1S)$ the melting occurs for a screening mass ${\mu}_m \simeq 1.50 GeV$. If we use eq.(7) this corresponds to $T_m\simeq 375MeV$. 

\vspace{0.2cm}

The corresponding times are also of significance:

\vspace{0.2cm}

a) $t_0$ is the time of formation of the plasma.

b) $t_c$ is the time at which the plasma cools below $T_c$ and becomes a hot hadronic gas.

c) $t_m$ is the time at which the plasma cools below the melting temperature $T_m$

\vspace{0.3cm}

These times acquire their significance when compared to the formation
time $t_f$, the time it takes for the initial $Q\bar Q$ pair, at rest,
to form a quarkonium bound state.  A simple model for $t_f$ \cite{22}
is the time required for the two quarks, forming a  specific
quarkonium state, to separate a distance R equal to the size of the
corresponding state.  The velocity of separation $v$ is taken to be
the average radial velocity of the quark as given by the potential model for that state, so $t_f=R/v$. 

The formation time $t_f$ increases with the temperature $T$, since $R$
increases and $v$ decreases when $T$ increases ( the screening mass
$\mu$ increases and the inter-quark potential weakens).  
We will be interested in what follows in $t_{f0}$ and $t_{fc}$, the
formation times at $T=0$ and $T=T_c$, respectively (note that $t_{f0}<t_{fc}$).   

\vspace{0.2cm}

We are now ready to discuss the fate of heavy $Q \bar Q$ bound states.

\vspace{0.3cm}

{\bf A}. For $T_c<T_m$: 

\vspace{0.2cm}

1. $t_c<t_{f0}$: the quark-antiquark pair escapes from the plasma
and forms a normal, $T=0$, cold state within the hadronic gas.

2. $t_{f0}<t_c<t_{fc}$: the pair does not reach the minimum size necessary
to form a hot bound state and the quarks are too far apart to form the
cold state in question when the QGP is replaced by the hadronic
gas. The state melts, although under some limited circumstance
an excited, cold resonance can form (for instance, $\Upsilon(2S)$
instead of $\Upsilon(1S)$).

3. $t_{fc}<t_c$: the pair forms a bound state in the plasma, a hot state. 
Its properties (e.g. mass and wave function) will be determined by the screened potential appropriate for some T, with $T_m>T>T_c$.

\vspace{0.3cm}

{\bf B}. For $T_c>T_m$ the picture is simpler: 

\vspace{0.2cm}

1. $t_c<t_{f0}$: the pair forms as a cold quarkonium state; 

2. $t_c>t_{f0}$: the quarkonium state cannot form in the plasma (too
hot) and it  melts or may form a cold, higher resonance.

\vspace{0.3cm}

Thus, for $Q \bar Q$ at rest, the formation of hot quarkonium states is
determined by the conditions:

\begin{equation}  
t_{fc}<t_c \quad and \quad T_c<T_m
\end{equation}  

\vspace{0.2cm}

Most $Q\bar Q$ are, however, not produced at rest with respect to the QGP rest frame. They will have transverse momentum ${\bf P_T}$ and we must account for the relativistic time dilation when computing the relevant $t_f$. The formation time in the QGP rest frame for a system of mass M and moving with momentum ${\bf P_T}$ is

\begin{equation}
{t'}_f({\bf P_T})=t_f\sqrt {1+{{\bf P_T}^2\over {M}^2}}	
\end{equation}

Hence eq.(9) is replaced by    

\begin{equation}
{t'}_{fc}({\bf P_T})<t_c,  \qquad T_c<T_m
\end{equation}

In order to relate the times $t_m$ and $t_c$ to temperatures we need an explicit expression for the entropies appearing in Eq.(8).  As usual we take the simplest reasonable form, i.e. the entropy for a free gas:

\begin{equation}
s(T)=const\cdot {{T}^3}
\end{equation}

Combining Eqs.(8),(9),(10),(11) and (12), we find that those $Q\bar Q$ pairs  
produced with transverse momentum in the range 
\begin{equation}
{\Bigl({{\bf P_T}\over M}\Bigr)}^2<{\Bigl({t_0\over t_{fc}}\Bigr)}^2{\Bigl({T_0\over T_c}\Bigr)}^6-1 
\end{equation}
will form hot quarkonium states.

\vspace{0.2cm}

$\bf{3.2 \quad Hot \quad \Upsilon \quad States \quad at \quad RHIC}$

\vspace{0.2cm}

Let us use eq.(13) for  $\Upsilon (1S)$. This state melts at ${\mu}_m\simeq1.50GeV$ and the corresponding melting temperature  
is $T_m=375MeV$, according to our choice $\mu(T)=4 \cdot T$. The formation
times needed are $t_{f0}=0.76fm$ and $t_{fc}=1.32fm$.

We will use values for $t_0$ and $T_0$ based on the expectations
at RHIC. Recent estimates \cite{21} suggest the QGP plasma should form at
$t_0\simeq 0.17fm$ with an initial energy density $\epsilon_{0}\simeq
98.2GeV/fm^{3}$ and initial temperature  $T_0=620MeV$. 

Using these values together with $M=9.615GeV$ in eq.(13), we
find that $\Upsilon (1S)$ states with  $P_{T}$ in the range

\begin{equation}
0<P_T<49Gev
\end{equation}
will be produced hot and have a mass shift of about
$160MeV$. Thus, according to our model, essentially all $\Upsilon(1S)$ produced
at RHIC will be ``hot''! The question we now ask is, will they survive?

\vspace{0.4cm}

$\bf{3.3 \quad Survival}$

\vspace{0.3cm}

If hot $\Upsilon$'s form in a QGP, as outlined above, they will be interesting only if we know of their existence.  The most promising signal of this existence is if some of the Upsilons  escape from the plasma and its surrounding fireball.  Since the hot $\Upsilon$ is still small, a geometric picture of scattering would imply a reasonable likelihood of escape.  

We picture a large, spherical plasma wich cools down to a hot hadron gas at 
time $t_c$. At this time all the hot $\Upsilon$s will be immersed in the 
hadron gas where they can scatter and cool down or be disrupted.  We
will quantify this in a simple absorption model, following the
treatment used by Karsch and Satz \cite{22} for the $J/\Psi$
absorbtion scenario in an expanding, hot hadronic gas.

The survival probability for the $\Upsilon$ entering the hadron gas at
time $t_c$ and surviving until time $\tau$ is

\begin{equation}
 S(\tau)=exp\Bigl[-\int_{t_c}^{\tau} dtn(t)\sigma\ \Bigr] 
\end{equation}                                                             

Here n(t) is the density of the medium, $\sigma$ is the appropriate hadronic
cross section for $\Upsilon(1S)$ and $\tau$ is the time it takes for the
hot hadron gas to become so cool and dilute that no further
interactions are likely. We take this 'freeze out' conditions to occur
when the energy density is $\epsilon (\tau)\simeq
0.3GeV/{fm^3}$ \cite{22}. If we continue to assume isentropic
longitudinal expansion for the hadron gas (treated as an ideal gas) as
well, the relation between densities at the relevant times is

\begin{equation}
 n(\tau)\tau=n(t_c)t_c(=n(t_0)t_0)
\end{equation}
and the survival probability becomes

\begin{equation}
 S(\tau)={ \Bigl[{n(\tau) \over n(t_c)} \Big]}^\kappa,\qquad \kappa=n_ct_c\sigma=n_0t_0\sigma
\end{equation}

As a final step, we use the relation between the density $n$ and the energy
density $\epsilon$ for an ideal gas, $n=({2\over 3}\epsilon
)^{3/4}$. This allows us to express $S$ in terms of the ``accessible''
quantities $\epsilon_0$,$t_0$ and $\sigma$:

\begin{equation}
S=\Bigl[{{\epsilon(\tau)}/{\epsilon}_c} \Big]^{(3/4) \cdot (2/3)^{3/4}\cdot {\epsilon}_0^{3/4}\cdot t_0 \cdot \sigma}
\end{equation}

Using $T_c=180MeV$ in the energy density for an ideal gas of gluons
and 3 flavours of massless quarks 

we find $\epsilon_c\simeq 2.14GeV$/${fm}^3$. Then, using the above-mentioned  value $\epsilon(\tau)\simeq 0.3GeV/{fm^3}$, the survival probability is:

\begin{equation}
S(\tau )=[0.14]^{(3/4)\cdot {(2/3)^{3/4}\cdot {\epsilon}_0}^{3/4}\cdot t_0\cdot \sigma} 
\end{equation}

The crucial unknown in this equation is the cross section for a hot
$\Upsilon $ interacting with hadrons.The escaping hot $\Upsilon (1S)$ can suffer two types of absorbtion in passing through the hot hadronic matter:

\vspace{0.2cm}

a) It can be dissociated by hadronic collisions, leading to supression
of the $\Upsilon$ signal similar to that initially observed for
$J/\Psi$. The survival probability is given by eq.(20) with $\sigma$
the inelastic cross section. 

In the limit of large quark mass, QCD allows us to calcculate [23-26] the inelastic scattering cross sections of $J/\Psi$ and
$\Upsilon$ on light hadrons (pions or nucleons). The
$asymptotic$ cross sections are estimated to be $\sigma
_{\Psi\pi}=(2-3)mb$ and $\sigma _{\Upsilon
\pi}=(0.6-0.8)mb$, which are comparable to the geometric cross
section. However, the asymptotic cross section is approached very
slowly because of the presence of a strongly damped threshold effect [23-25]. For instance, for $E_{\Upsilon}<100GeV$, $\sigma_{\Upsilon
\pi}<0.1mb$. The corresponding predictions for $J/\Psi$ inelastic
scattering have been succesfully tested against photo-production cross
sections [27]. 

Therefore we adopt a (conservative) estimate of $0.3mb$ for the inelastic
(dissociation) cross section of the hot $\Upsilon (1S) $.  

Using this value in eq.(19) in conjunction with ${\epsilon}_0=98.2GeV$/${fm}^3$ and $t_0=0.17fm$ \cite{21}, we find   $S \simeq 0.56$ i.e. $56\%$ of the produced $\Upsilon (1S)$ still survive the hadronic fireball.

\vspace{0.2cm}

b) $\Upsilon (1S)$ can scatter quasi-elastically into the normal, $T=0$, $\Upsilon (1S)$. This is the process that is of most interest to us, since it decides wheather or not a significant number  of hot  $\Upsilon (1S)$ could be detected. 

To make an estimate to the fraction of hot $\Upsilon (1S)$ versus cold
($T=0$) $\Upsilon (1S)$ we therefore need to know the cross section
$\sigma_{qe}$ associated to the reaction

\begin{equation}
\Upsilon (1S)(hot)+hadron \rightarrow \Upsilon (1S)(cold) +hadron'
\end{equation}

The QCD approach to heavy quarkonium scattering can also be applied to
the elastic and quasi-elastic scattering \cite{23,26}. These cross
sections are expected to be quite small, considerably smaller
than the inelastic cross sections, an expectation that is again born
out by analysis of the photo-production data [27].

To be conservative we use a value $\sigma=0.1mb$ in eq.(19) and obtain $S\simeq 0.82$, i.e.  $82\%$ of produced hot $\Upsilon (1S)$ states will survive their traversal of the hot hadron gas as hot, mass shifted states.

\vspace{0.2cm}

To conclude this section, let us review the results:                           

\vspace{0.1cm}

i) all directly produced $\Upsilon (1S)$ states with transverse momenta below $49GeV$ are hot, with a mass shift of over $150MeV$; 

\vspace{0.1cm}

ii) $56\%$ of them will survive the passage through the hadronic
aftermath either as hot or cold states;

\vspace{0.1cm}

iii) the fraction of surviving $\Upsilon (1S)$ which are superheated
is (at least!) $82\%$.

\vspace{0.2cm}

$\bf{3.4\quad Sensitivity \quad to \quad input \quad parameters}$

\vspace{0.2cm}

Our result is sensitive to the values we have assigned for the initial
energy density ${\epsilon}_0(=98.2GeV/{fm}^3$), the hadronization temperature
$T_c=180MeV$ (wich determines ${\epsilon}_c$), the formation time of
the plasma
$t_0 (=0.17fm$) and the quasi-elastic cross section $\sigma (=0.1mb)$,
none of wich are precisely known. 
Our key result, the fraction $S$ of surviving $\Upsilon $'s 
that are superheated, changes as follows when we vary these values one at a time (within reasonable limits):

\vspace{0.1cm}

a) ${\epsilon}_0=80GeV/{fm}^3$ gives $S=0.85$ and ${\epsilon}_0=120GeV/{fm}^3$ gives $S=0.80$

\vspace{0.1cm} 

b) $T_c=170MeV$ gives  $S=0.84$ and $T_c=190MeV$ gives $S=0.81$

\vspace{0.1cm}

c) Larger value of $\sigma =0.3mb$ gives $S=0.56$ 

\vspace{0.1cm}

d) varying $t_0$ affects twofold our result for the fraction of hot
$\Upsilon $s available for detection. Firstly, $t_0$ enters  eq.(13)
which determines the maximum $P_T$ for hot $\Upsilon $s (according to eq.(13), $t_0$ has to be larger than $0.032fm$ in order to have hot $\Upsilon $'s formed).  Secondly, $t_0$ enters eq.(19), affecting the value of $S$. Values of $t_0$ from $0.1fm$ to $0.25fm$ ensure values of values of $P_{Tmax}$ between $28GeV$ and $73GeV$ and values of $S$ between $89\%$ and $75\%$.
 
\vspace{0.2cm}

This range of parameters is probably sufficiently broad that it might
encompass changes in the underlying models we have used to describe
our effect.

\vspace{0.3cm}

$\bf{3.5 \quad Contribution \quad from \quad Higher \quad States}$

\vspace{0.3cm}                                                                            

Our discussion dealt so far with directly produced $\Upsilon (1S)$ states. We look now into the contribution from the excited states. 

In Table 2 we present the formation times (at $T=0$) and melting $\mu$ values for the higher bottomonium states that contribute significantly to the $\Upsilon (1S)$ production by cascade decays.

\vspace{0.2cm}

\begin{table}[h]
\begin{tabular}{|c|c|c|c|c|}
\hline
 & $\Upsilon (2S)$ &
$\Upsilon (3S)$ & $\chi (1P)$ &
$\chi (2P)$ \\\hline
radius \quad at \quad T=0 \quad (fm)        & 0.51          & 0.75           & 0.41      & 0.66      \\\hline 
formation \quad time \quad $t_{f0}$(fm)             & 1.81          & 2.46           & 2.50      & 2.88      \\\hline 
melting \quad $\mu$ \quad (GeV)               & 0.63          & 0.42
& 0.58      & 0.41      \\\hline
$T_m$\quad [from\quad eq.(7)]\quad (GeV)      & 0.16          & 0.11 
&0.15     & 0.10
\\\hline    
\end{tabular}
\caption{Bound state properties of excited $\Upsilon$ states}
\end{table}
 
\vspace{0.2cm}

We notice that the melting temperatures, according to eq.(7), are below $T_c$ and therefore these states cannot form in the QGP. Their formation has to take place in the post-plasma era (that is, within a hadronic medium) and consequently their binding properties  should be the normal ones (there are no hot $\chi_b$'s, e.g.).

Recalling the analysis of $\bf{3.1}$, we determine the minimum
transverse momentum these states should have in order to escape
melting within the QGP at RHIC ($T_0=620MeV$). 

\vspace{0.3cm}

\begin{table}[h]
\begin{tabular}{|c|c|c|c|c|}
\hline
& $\Upsilon (2S)$ & $\Upsilon (3S)$  & $\chi_{b}(1P)$ & $\chi_{b} (2P)$ \\\hline
$P_{Tmin}(GeV)$       & 37            & 27.3             & 25.7            & 22.5        \\\hline 
\end{tabular}
\caption{Minimum tranverse momenta to escape melting in QGP (RHIC)}
\end{table}

We find that the feed-down production of $\Upsilon(1S)$ from higher
states contributes only for transverse momenta larger than the values
shown in  Table 3. Consequently, we expect that at RHIC $all$
$\Upsilon(1S)$ with $P_T$ below $22.5GeV$ are directly produced. 

\vspace{0.3cm}

$\bf{3.5 \quad Discussion \quad on \quad functional\quad dependence\quad
of \quad \mu(T)}$

\vspace{0.3cm}

The numerical results presented so far assume the validity of
eq.(7). Changes in functional dependence $\mu(T)$ generate changes in
the parameters $t_{fc}$ (the formation time at $T=T_c$) and $T_m$ (the melting temperature). According to the analysis presented in ${\bf 3.1}$
we find the following results when the functional dependence $\mu(T)$
is allowed to vary from $2.5 \cdot T$ to
$6 \cdot T$:

\vspace{0.2cm}

i) For $\mu = 2.5 \cdot T$, all $\Upsilon(1S)$ with
transverse momenta $0<P_T<66GeV$ are formed as hot states. 
The only change regarding the higher states is that now there are hot
$\Upsilon(2S)$ formed with transverse momenta $0<P_T<5.8GeV$
that will eventually contribute to (cold) $\Upsilon(1S)$ signal. 

This possibility will disappear at  $\mu \simeq 2.8 \cdot T$ and for
$\mu/T = 2.8-6$ there is essentially no change, i.e. the conclusions of
${\bf{3.4}}$ are entirely valid, as far as the contribution
from the higher states are concerned.

\vspace{0.2cm}

ii) When $\mu/T$ increases to $6$, the momentum range of hot $\Upsilon(1S)$
states decreases to  $0<P_T<24.8GeV$. 

\vspace{0.2cm}

Note that in the (simple) model we have used to find the survival
probability of a hot $\Upsilon(1S)$ state, the function $S$ (see eq.(19))
has no $P_T$ dependence. Thus there are no changes in $S$ when $\mu
(T)$ is varied as mentioned above.

We then see that despite the large range allowed for the $\mu(T)$
functional form, a significant number of hot $\Upsilon(1S)$ states could be
produced at RHIC with negligible feed-down contribution from
higher resonances.

\section{\bf Conclusions}

We conclude that there is a reasonable possibility that hot
$\Upsilon(1S)$ states, with masses measurably different from the $T=0$
value will be produced, escape and be detected. More specifically, our
prediction is that about $80\%$ of all $\Upsilon(1S)$ states detected
at small rapidity and with transverse momenta below $25GeV$ will have a mass shift of $\sim 150MeV$, this shift being caused by their formation in the screening QGP.

The arguments we have presented are partly qualitative, similar to the original arguments in favor of $J/\Psi$ suppression. Nevertheless the strong physical pictures that underlies these arguments leads us to propose a search for superheated $\Upsilon $s as a signal for Quark Gluon Plasma formation.

\section{\bf Acknowledgment}

We are indebted to our colleagues for useful discussions. This work was supported in part by the U.S.Department of Energy.


\begin{thebibliography}{99}

\bibitem{1}T. Matsui, H. Satz: Phys. Lett. 178B (1986), 416 
\bibitem{2}R. Vogt: Phys. Rep. 310 (1999), 197; H. Satz: arXiv:hep-ph/0007069 
\bibitem{3}M.C. Abreu et al.(NA50 Collaboration): Phys. Lett. 450B (1999), 456;
Phys. Lett. 410B (1997),327;Phys. Lett. 410B (1997), 337
\bibitem{4}D. Kharzeev, M. Nardi, K. Satz:arXiv:hep-ph/9707308
\bibitem{5}M. Nardi, K. Satz: Phys. Lett. 442B (1998), 14
\bibitem{6}E. Shuryak, G. Teyner: arXiv:nucl-th/9801016
\bibitem{7}C. Spieles et al., Phys. Rev. C 60 (1999), 054901-1
\bibitem{8}N. Armesto, A. Capella, E.G. Ferreiro: Phys. Rev. C 59(1999), 395
\bibitem{9}W. Cassing, E.L. Bratkovskaya: Nucl. Phys. A 623 (1997), 570
\bibitem{10}J.F. Gunion, R. Vogt: Nucl. Phys. B 492 (1997), 301
\bibitem{11}F. Karsch, M.T. Mehr, H. Satz: Z. Phys. C- Particles and Fields 37 (1988), 617
\bibitem{12}D.B. Lichtenberg, E. Predazzi, R. Roncaglia, M. Rosso, J.G. Wills: Z. Phys. C41 (1989), 615 

\bibitem{13}D.J. Gross, R.D. Pisarski, L.G. Yaffe: Reviews of Modern
Physics 53 (1981), 43
\bibitem{14}U.M. Heller,F. Karsch, J. Rank: Phys. Rev. D57 (1998), 1438
\bibitem{15}S. Datta, S. Gupta: Phys. Lett. B471 (2000), 382
\bibitem{16}S. Datta, S. Gupta: Nucl. Phys. B534 (1998), 392
\bibitem{17}A. Hart, M. Laine, O. Philipsen: arXiv:hep-ph/0004060, to
appear in Nucl. Phys. B
\bibitem{18}K. Kajantie et al. : Phys. Rev. Lett. 79 (1997), 3130
\bibitem{19}J.D. Bjorken: Phys.Rev. D 27 (1983), 140
\bibitem{20}F. Karsch: arXiv:hep-lat/9909006
\bibitem{21}K.J. Eskola et al. : Nucl. Phys. B570 (2000), 379
\bibitem{22}F. Karsch,H. Satz: Z. Phys. C- Particles and Fields 51 (1991), 209 

\bibitem{23}G. Bhanot, M. Peskin: Nucl. Phys. B 156 (1979), 391
\bibitem{24}A. Kaidalov: Proceedings,XXVIII Rencontres de Moriond
(1993), ed. J. Tran Thanh Van, pg. 601
\bibitem{25}D. Kharzeev, K. Satz: Phys. Lett. 334B (1994), 155

\bibitem{26}H. Fujii, D. Kharzeev: Phys. Rev. D60 (1999), 114039 
\bibitem{27}K. Redlich, H. Satz, G.M. Zinovjev: arXiv:hep-ph/0003079


\end{thebibliography}
\end{document}